\documentclass[prl,twocolumn,amsmath,amssymb,amsfonts,superscriptaddress,]{revtex4-1}

\usepackage{graphicx,xcolor,mathtools,bm,braket,newtxtext,newtxmath,chemformula}
\usepackage[hyphenbreaks]{breakurl}
\usepackage[normalem]{ulem}
\usepackage[colorlinks=true,linkcolor=blue,citecolor=blue,urlcolor=blue]{hyperref}

\newcommand{\normord}[1]{\vcentcolon\mathrel{#1}\vcentcolon}
\providecommand{\vcentcolon}{\mathrel{\mathop{:}}}
\newcommand{\Tr}{\mathop{\mathrm{Tr}} \nolimits}

\begin{document}
	
\title{Experimental Demonstration of the Field Commutator Using Intensity Correlations}
	
\author{Hans Dang}
\affiliation{Department Physik, Friedrich-Alexander-Universit\"at Erlangen-N\"{u}rnberg, 91058 Erlangen, Germany}
\affiliation{Max-Planck-Institut f\"{u}r die Physik des Lichts, 91058~Erlangen, Germany} 
	
\author{Sebastian Luff}
\affiliation{Department Physik, Friedrich-Alexander-Universit\"at Erlangen-N\"{u}rnberg, 91058 Erlangen, Germany}
\affiliation{Max-Planck-Institut f\"{u}r die Physik des Lichts, 91058~Erlangen, Germany} 
	
\author{Martin Fischer}
\affiliation{Max-Planck-Institut f\"{u}r die Physik des Lichts, 91058~Erlangen, Germany} 
	
\author{Markus Sondermann}
\affiliation{Department Physik, Friedrich-Alexander-Universit\"at Erlangen-N\"{u}rnberg, 91058 Erlangen, Germany}
\affiliation{Max-Planck-Institut f\"{u}r die Physik des Lichts, 91058~Erlangen, Germany}

\author{Mojdeh.~S.~Najafabadi}
\affiliation{Max-Planck-Institut f\"{u}r die Physik des Lichts, 91058~Erlangen, Germany}
	
\author{Luis~L.~S\'anchez-Soto}
\affiliation{Max-Planck-Institut f\"{u}r die Physik des Lichts, 91058~Erlangen, Germany}
\affiliation{Departamento de \'Optica, Facultad de F\'{\i}sica, Universidad Complutense, 28040~Madrid, Spain}
\affiliation{Institute for Quantum Studies, Chapman University, Orange, CA 92866, USA}
	
\author{Gerd~Leuchs}
\affiliation{Department Physik, Friedrich-Alexander-Universit\"at Erlangen-N\"{u}rnberg, 91058 Erlangen, Germany}
\affiliation{Max-Planck-Institut f\"{u}r die Physik des Lichts, 91058~Erlangen, Germany}
\affiliation{Department of Physics, University of Ottawa, Ottawa, {Ontario K1N 6N5}, Canada}
	
\begin{abstract}
The canonical commutation relation is a cornerstone of quantum theory and underlies the Heisenberg uncertainty principle. Although uncertainty relations have been extensively tested, direct verifications of the underlying commutation relation itself have remained elusive. Here, we report an experimental demonstration of the bosonic commutation relation for optical field operators based on measurements of two distinct intensity correlation functions. From these measurements, we extract the expectation value of the field-operator commutator for both a single-photon state and a coherent state. In each case, the measured commutator is consistent with unity, in precise quantitative agreement with quantum theory, demonstrating a quantum effect independent of the specific statistical properties of the light field. 
\end{abstract}
	
\maketitle
	
\textit{Introduction.---}
Heisenberg’s uncertainty principle, whether framed in terms of indeterminacy~\cite{Heisenberg:1927} or complementarity~\cite{Bohr:1928}, is rooted in a fundamental feature of quantum mechanics: the noncommutativity of operators associated with complementary observables. In stark contrast to classical intuition, this principle asserts that certain pairs of physical properties, such as position and momentum or wave-like and particle-like behavior, cannot be simultaneously specified with arbitrary precision in the same experiment. This inherent uncertainty is not a limitation of measurement, but a fundamental property of nature itself.
	
Although the uncertainty principle has been rigorously tested across a wide range of physical systems, direct experimental access to the commutation relations that underpin it remains elusive: noncommuting observables admit no joint probability distribution, and their product is generally non-Hermitian, precluding straightforward measurement. Demonstrations to date have been largely confined to finite-dimensional qubit systems~\cite{Hasegawa:1997,Wagh:1997,Kim:2010,Yao:2010,Wagner:2021}. In the continuous-variable domain, the canonical commutation relation between position and momentum—or, equivalently, between bosonic creation and annihilation operators—has been probed only indirectly, via sequential single-photon addition and subtraction from an optical field~\cite{Zavatta:2004,Zavatta:2009}.
	
We address this question from a different perspective by exploiting intensity correlation measurements, a technique pioneered by Hanbury Brown and Twiss (HBT)~\cite{Hanbury-Brown:1956a,Hanbury-Brown:1956b} and formally grounded in Glauber’s quantum theory of optical coherence~\cite{Glauber:1963a,Glauber:1963b,Glauber:1963c}. In the conventional HBT experiment, an optical field is incident on a 50:50 beam splitter, and the two output fields are detected by independent detectors whose signals are subsequently time-correlated. This scheme enabled the landmark observation of photon antibunching~\cite{Kimble:1977a,Cresser:1982a,Walls:1979a, Leuchs:1986a}, which is long regarded as compelling  evidence for the particle nature of light~\cite{Grangier:1986a}.
	
Yet the HBT arrangement, though the gold standard for measuring intensity correlations, can misleadingly suggest that a beam splitter is essential to reveal photon indivisibility. As Loudon pointed out~\cite{Loudon:1973aa}, a far simpler configuration,  a single detector illuminated directly, suffices in principle, with correlations extracted from the time-resolved detection record via autocorrelation analysis~\cite{Wiersig:2009,Steudle:2012}. This method is computationally efficient but yields a pronounced peak at zero delay, absent from cross-detector measurements, as expected for any autocorrelation function. One might attribute this peak to independent, delta-correlated Poissonian noise intrinsic to each detector, but this would require every detector to possess such noise, a realistic assumption only for detectors with vanishing efficiency. For any other detector, including those used here, no classical stochastic theory can account for it.
	
In this Letter, we demonstrate that the distinction between such cross- and auto-correlation measurements is not merely technical, but reflects a fundamental consequence of the noncommutativity of bosonic field operators. Specifically, HBT cross-correlations probe a normal ordering of field operators that is intrinsically different from the operator ordering accessed in single-detector auto-correlations. By quantitatively comparing these two classes of intensity correlations, we show that the zero-delay auto-correlation peak—commonly regarded as the variance of the detector noise—constitutes a direct experimental manifestation of the bosonic commutation relation. 
	
We further generalize this approach to establish a framework for determining commutation relations through intensity correlation measurements. Remarkably, our method applies to arbitrary optical fields and operates independently of whether their statistics are 'classical' or 'quantum', underscoring both its fundamental significance and its broad experimental accessibility.

\textit{Theoretical background.---}
We consider the standard HBT setup. Within Glauber's photodetection theory~\cite{Glauber:1963b}, the joint probability of registering one photon at time $t$ in the first detector and a second photon at time $t^{\prime}$ in the second detector is proportional to $w(t, t^{\prime}) \propto \langle \hat{E}^{(-)} (t + \tau) \hat{E}^{(-)} (t) \hat{E}^{(+)} (t) \hat{E}^{(+)} (t + \tau) \rangle := G^{(2)} (t,t^{\prime} )$, where $\tau = t^{\prime} - t$ is the delay, $\hat{E}^{(\pm)}$ are the positive- and negative-frequency parts of the field, and the angle brackets denote ensemble averages taken with the density matrix, $\langle \hat{O} \rangle = \Tr (\hat{\varrho} \hat{O})$. Throughout, we take the field to be ergodic, so that ensemble averages coincide with time averages~\cite{Mandel:1995a}.

The measurement therefore returns the normalized second-order correlation function
\begin{equation}
	g^{(2)}_{\mathrm{cross}}( \tau) = \frac{ \langle \normord{\hat{I} (t) \, \hat{I}(t+\tau)} \rangle}{\langle {I(t)} \rangle\langle {I(t +\tau )} \rangle} = \frac{\langle \hat{a}^\dagger (t + \tau)\hat{a}^\dagger (t) \hat{a}(t)\hat{a}(t+\tau) \rangle}{\langle \hat{a}^\dagger (t) \hat{a}(t) \rangle\langle \hat{a}^\dagger (t + \tau) \hat{a}(t + \tau) \rangle} \, ,
	\label{normal ordering prob}
\end{equation}
where $\hat{I} (t)= \hat{E}^{(-)} (t) \hat{E}^{(+)} (t)$ is the intensity operator and the two colons $\normord{ \, }$ denote normal ordering. The subscript ``cross'' stresses that we cross-correlate the signals measured with two different detectors, including the correlation at zero delay $\tau=0$.   For stationary fields the definition does not depend on the absolute detection time $t$ and the two terms in the denominator are equal.

A key feature of $g_{\mathrm{cross}}^{(2)}(\tau)$ is its independence from optical losses, which makes it particularly robust experimentally. This robustness, however, comes at the expense of reduced sensitivity to absolute photon-number statistics. In addition, the HBT configuration is largely unaffected by detector dead time. All of this applies to energy-resolving detectors, such as photodiodes or click detectors, but not to quadrature-sensitive schemes such as homodyne detection~\cite{Bjork:2001a,Hessmo:2004a,Shikhali:2024aa,bozyigit2011}.

Most intensity-correlation measurements employ the HBT scheme. One may nevertheless ask why the correlation function is not instead obtained by recording a time series with a single detector and evaluating the correlation numerically~\cite{Steudle:2012}. We denote the resulting auto-correlation function as  $g^{(2)}_{\mathrm{auto}}$. Unlike the HBT cross-correlation, it  does generally depend on losses, a consequence of the nontrivial effect of attenuation on quantum signals. Importantly, however, losses affect only the zero-delay value: for all other values $\tau \neq 0$, the two correlation functions are identical, as outlined below.

This distinction has practical value. Rather than splitting the incoming light on a beam splitter and cross-correlating two time series, it may suffice to send all of it to a single detector with a fast response and auto-correlate the result.  Depending on the delay range of interest, this can significantly reduce the required measurement time. For example, sending $N$ photons into an HBT setup with a beam splitter of reflectivity $R$ and transmissivity $1 - R$, the expected number of correlations scales as $NR \times N(1-R) = N^2(R-R^2)$. In contrast, when the same photons are detected by a single detector and numerically auto-correlated, the number of correlations at $\tau \neq 0$ scales as $N^2$. Hence, the number of data points contributing to $g^{(2)}_{\mathrm{cross}}(\tau )$ scales as $\mathcal{O}\left(N^2(R-R^2)\right)$, whereas for $g^{(2)}_{\mathrm{auto}}(\tau)$ it scales as $\mathcal{O} (N^2)$. For the optimal case of a 50:50 beam splitter, this translates to a reduction in the required measurement time by a factor of four.

The two approaches differ, however, at zero delay. The discrepancy arises because $g^{(2)}_{\mathrm{cross}}(\tau)$ involves the product of two independent signals measured by different detectors, whereas $g^{(2)}_{\mathrm{auto}}(\tau)$ corresponds to numerically correlating a signal with itself. For the split fields, these operations are not equivalent due to the projection associated with quantum measurement~\cite{Khan:2017a}. In particular, attenuating a train of single-photon pulses bythe beam splitter does not result in uniformly reduced pulse amplitudes; instead, some pulses are removed entirely while others remain unchanged. A photon thus appears in a given time bin only with a small probability, but when it does, it is detected as a full photon. Auto-correlating such a signal therefore assigns unit conditional detection probability to a second photon at $\tau=0$, rather than a reduced probability reflecting optical losses.

\begin{figure}[t]
	\centering
	\includegraphics[width=\linewidth]{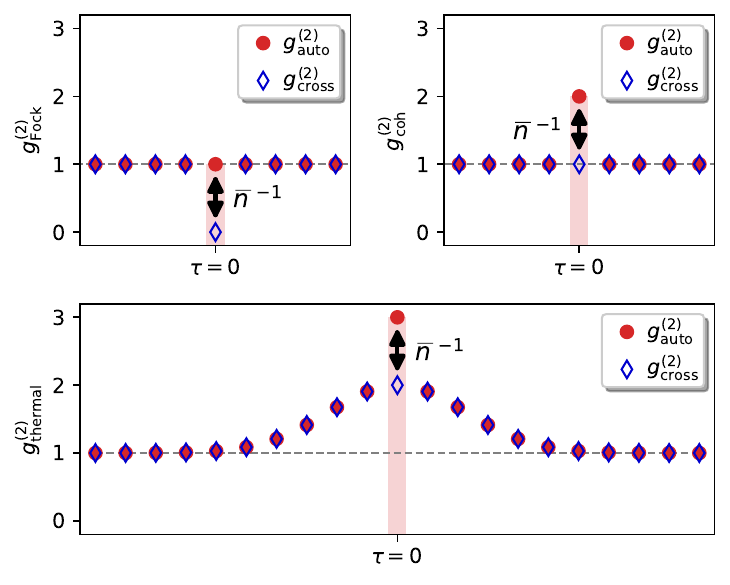}
	\caption{Comparison of theoretical curves for $g^{(2)}_{\mathrm{auto}}(\tau)$ and $g^{(2)}_{\mathrm{cross}}(\tau)$ for a time bin width equal to the inverse detector bandwidth. Top left: perfect sequence of single photon states with one photon per time bin; $g^{(2)}_{\mathrm{cross}}(\tau)$ shows anti-bunching. Top right: coherent state. Bottom: thermal light state. In all cases the difference between the normalized auto and cross correlation functions is equal to the inverse of the detected mean number of photons $\langle n\rangle =\langle \hat{a}^{\dagger} (t) \hat{a} (t)\rangle $.} 
	\label{comparison}
\end{figure}

Quantum theory captures this effect nicely. Recording a single intensity time series and correlating it with itself—typically averaging over time—leads to the auto-correlation function
\begin{equation}
g^{(2)}_{\mathrm{auto}}(\tau) 
= \frac{\langle \hat{a}^\dagger (t) \hat{a}(t+\tau)\hat{a}^\dagger (t) \hat{a}(t+\tau) \rangle}{\langle \hat{a}^\dagger (t) \hat{a}(t) \rangle\langle \hat{a}^\dagger (t+\tau) \hat{a}(t+\tau) \rangle} \, .
\label{g2-auto1}
\end{equation}
For a pure quantum state $|\psi \rangle$, the difference between the auto- and cross-correlation functions at zero delay is given by
\begin{equation}
	g^{(2)}_{\mathrm{auto}}(0) - g^{(2)}_{\mathrm{cross}}(0) = \frac{
	\langle \phi | [\hat{a} (t), \hat{a}^{\dagger} (t) ] | \phi \rangle}{ \langle \hat{a}^{\dagger} (t) \hat{a} (t) \rangle \, } \, , 
	\label{eq:ccr}
\end{equation}
where $|\phi \rangle = 1/\sqrt{\langle \hat{a}^{\dagger} \hat{a} \rangle} \; \hat{a} |\psi \rangle$ is the state of the field immediately after a photon has been detected. Thus, the difference $g^{(2)}_{\mathrm{auto}}(0) - g^{(2)}_{\mathrm{cross}}(0) $ provides direct access to the bosonic commutation relation. In what follows, we exploit this relation to test these fundamental quantum properties. We note that for $\tau\neq 0$ no such simplified relationship can be made, unless the field operators at different times commute (see below).

Figure~\ref{comparison} shows the comparison of auto- and cross-correlation functions for three different light fields:  perfect \(n=1\) Fock state, coherent state  and thermal state. The bin width is matched to the inverse detector bandwidth, so that the difference is confined to a single bin, where it equals the inverse mean photon number. In the experimental curves below the signal is oversampled, with bins much narrower than the inverse bandwidth.

\textit{Auto- vs cross-correlation.---} 
To experimentally assess the difference $g^{(2)}_{\mathrm{auto}}(\tau) - g^{(2)}_{\mathrm{cross}}(\tau) $ we focus first on $\tau =0$ by using the commutator for the field operators. After integrating over time, this yields
\begin{equation}
	g^{(2)}_{\mathrm{auto}}(0) - g^{(2)}_{\mathrm{cross}}(0) = \frac{1}{\langle \hat{a}^\dagger (t) \hat{a}(t) \rangle}   ,
	\label{g2-auto2}
\end{equation}

In a realistic experimental setting, however, the number of detected photons $N_{\mathrm{det}}(t)$ generally differs from the ideal mean photon number $\langle \hat{a}^{\dagger} (t) \hat{a}(t) \rangle$. The two are related by
\begin{equation}
	\langle N_{\mathrm{det}} (t)\rangle = \frac{\min \{ A_{\mathrm{det}},A_{\mathrm{mode}}\}}{A_{\mathrm{mode}}} \, \frac{T_{\mathrm{bw}}}{T_{\mathrm{coh}}} \, \eta \, \langle \hat{a}^\dagger (t) \hat{a}(t) \rangle \, ,
\end{equation} 
where three multiplicative factors account for the experimental imperfections.

The first is the spatial overlap between the optical mode and the detector: whenever the mode cross-section $A_{\mathrm{mode}}$ exceeds the photosensitive area $A_{\mathrm{det}}$, the light falling outside that area is simply lost. The second is the temporal overlap. A coherence time $T_{\mathrm{coh}}$ longer than the measurement interval $T_{\mathrm{bw}}$ reduces the signal collected per bin; a shorter one enhances it. The third, $\eta$, collects all remaining optical losses, including the non-unit quantum efficiency of the detector itself.

Together these factors define an effective detection efficiency $\eta_{\mathrm{eff}}$. It is conveniently modeled by a fictitious beam splitter of transmission $\eta_{\mathrm{eff}}$ placed in front of an otherwise ideal detector: a fraction $\eta_{\mathrm{eff}}$ of the incident photons reaches the detector, and the rest escapes into an unobserved vacuum mode~\cite{Meda:2017aa}. The difference between the two correlation functions then becomes
\begin{equation}
g^{(2)}_{\mathrm{auto}}(0) - g^{(2)}_{\mathrm{cross}}(0) = 
 \frac{1}{ \eta_\mathrm{eff} \, \langle \hat{a}^\dagger (t) \hat{a}(t) \rangle} \,  ,
\label{g2-auto2correction}
\end{equation}
that is, the inverse of the mean number of photons actually detected.

For $\tau \neq 0$, a straightforward extension of the arguments leading to \eqref{eq:ccr}  yields
\begin{equation}
	g^{(2)}_{\mathrm{auto}}(\tau) - g^{(2)}_{\mathrm{cross}}(\tau) = \frac{
	\langle \phi (t)| [\hat{a} (t), \hat{a}^{\dagger} (t^{\prime}) ] | \phi (t^{\prime}) \rangle}{ \sqrt{\langle \hat{a}^{\dagger} (t) \hat{a} (t) \rangle \langle \hat{a}^{\dagger} (t^{\prime}) \hat{a} (t^{\prime}) \rangle}  \, } \, .
\end{equation}

\begin{figure}[t]
	\centering
	\includegraphics[width=\linewidth]{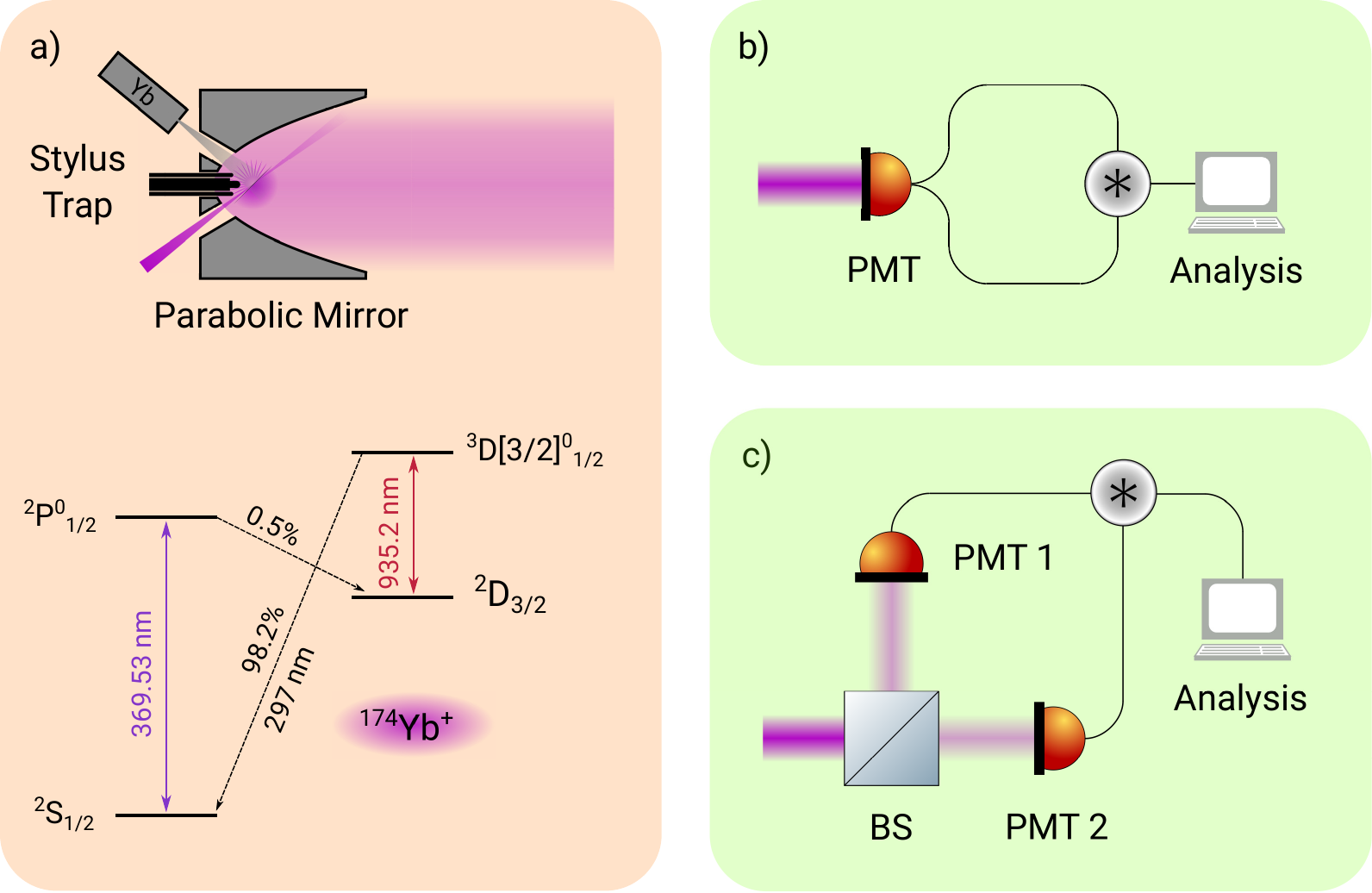}
	\caption{a) Sketch of the parabolic mirror with inserted trap assembly and \(^{174}\mathrm{Yb}^{+}\) energy levels. The light scattered towards the detection setup is either emitted by the ion at the focus of the parabolic mirror (ion-focus distance not to scale) or laser light scattered from one of the trap electrodes. b) and c) Sketches of the used detection methods.}
	\label{fig:setup}
\end{figure}

\textit{Experimental results.---}
We now translate these considerations into two experimental scenarios. In each, we measure $g^{(2)}_{\mathrm{auto}}(\tau)$ and $g^{(2)}_{\mathrm{cross}}(\tau)$ independently and then apply Eqs.~\eqref{eq:ccr} and~\eqref{g2-auto2correction}.

The first experiment uses a single-photon source: one $^{174}\mathrm{Yb}^{+}$ ion trapped at the focus of a deep parabolic mirror, in a setup similar to Ref.~\cite{Maiwald:2012a}. The ion is driven continuously by a 369.5~nm laser close to resonance on the $^{2}\mathrm{S}_{1/2}\leftrightarrow{}^{2}\mathrm{P}_{1/2}$ dipole transition, whose upper-state lifetime of $8.1$~ns~\cite{Berends:1993a} sets the coherence time $T_{\mathrm{coh}}$ of the emitted photons. The parabolic mirror collimates the fluorescence, and a pinhole in the Fourier plane of a telescope suppresses straylight from the driving laser. A $50:50$ beam splitter then divides the single photons between two photomultiplier tubes (PMTs), whose detection events are recorded with a time-to-digital converter and analyzed numerically (see Fig.~\ref{fig:setup}). A repump laser at 935~nm returns population that decays into the $^{2}\mathrm{D}_{3/2}$ state, closing the cooling cycle [see Fig.~\ref{fig:setup}(a)].

The detection efficiencies of the two PMTs in the setup were determined using the state sensitive method described in the Supplemental Material, obtaining $1.94 \pm 0.02~\%$ for the detector in transmission and $2.67 \pm 0.02~\%$ for the one in reflection, hence $\eta = 4.61 \pm 0.03~\%$ for the two combined.  

The same time-tag streams from both detectors were used to evaluate $g^{(2)}_{\mathrm{auto}}(\tau)$ and $g^{(2)}_{\mathrm{cross}}(\tau)$. For $g^{(2)}_{\mathrm{auto}}(\tau)$, the ``single detector'' was realized by first merging the two time-tag streams into a single stream and then correlating it with itself [see Fig.~\ref{fig:setup}(b)]. By merging the two streams, detector-specific artifacts such as photomultiplier dead time are effectively suppressed as in a standard HBT measurement. The cross-correlation $g^{(2)}_{\mathrm{cross}}(\tau)$ was obtained by correlating the time-tag stream from one detector with that of the other [see Fig.~\ref{fig:setup}(c)]. In both cases, correlations were calculated by counting coincidence events at time delays $\tau$ within a time window of $T_\text{bw}$, followed by appropriate normalization.

\begin{figure}
\includegraphics[width=\columnwidth]{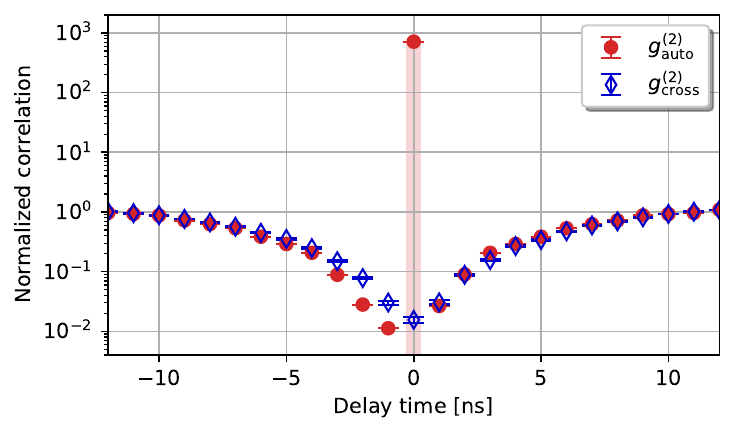}
\caption{Measured correlation functions $g^{(2)}_{\mathrm{auto}}$ and $g^{(2)}_{\mathrm{cross}}$ of a train of single photons emitted by a single trapped ion. Both functions are nearly identical and only differ for $\tau = 0$, as expected from theory. The characteristic antibunching dip for a single-photon source is also apparent from the correlation functions.}
\label{fig:single_photon_source_correlation}
\end{figure}

The resulting correlation functions for a train of single photons emitted by a single ion are shown in Fig.~\ref{fig:single_photon_source_correlation}, evaluated with a binwidth of $ T_{\mathrm{bw}} = 1$~ns. Both correlation functions are nearly identical, except at zero delay, in reasonable agreement with the theoretical prediction. The pronounced antibunching dip is clearly visible, confirming the single-photon character of the source. We obtain $g^{(2)}_{\mathrm{auto}}(0) = 714.93 \pm 0.19$ and $g^{(2)}_{\mathrm{cross}}(0) = 0.016 \pm 0.002$. In the ideal limit of vanishing straylight, $g^{(2)}_{\mathrm{cross}}(0)$ should be zero for a single-photon source.

Per excited-state lifetime, the ion emits on average $\beta \rho_{22}$ photons on the $^{2}\mathrm{S}_{1/2}\leftrightarrow{}^{2}\mathrm{P}_{1/2}$ transition, where $\rho_{22}$ is the excited-state population and $\beta$ the branching ratio into that transition.  Only a fraction $\eta_{\mathrm{eff}}$ of them are detected within the window $T_{\mathrm{bw}}$, so the mean number of detected photons per bin is $\langle N_{\mathrm{det}} \rangle = \beta \rho_{22} \eta_{\mathrm{eff}} T_{\mathrm{bw}} / T_{\mathrm{coh}} = (1.42 \pm 0.01) \times 10^{-3}$. Inserting the measured correlations into Eq.~\eqref{eq:ccr} then gives $\langle \phi \vert [\hat{a},\hat{a}^{\dagger}]\vert\phi\rangle = 1.01 \pm 0.04$, in excellent agreement with the expected value of unity.

\begin{figure}[t]
	\includegraphics[width=\columnwidth]{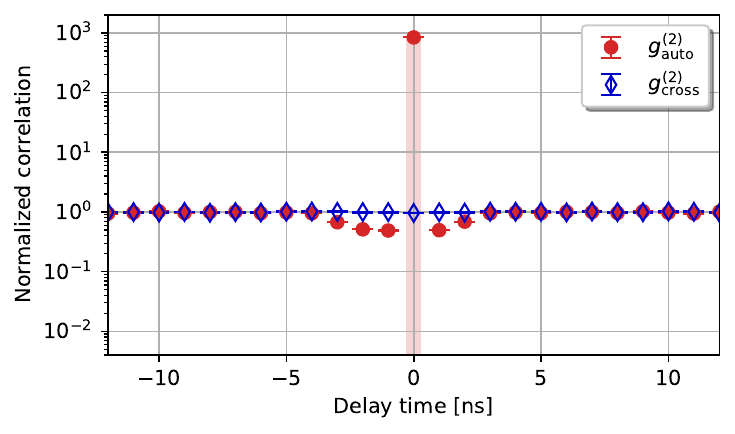}
	\caption{Measured correlation functions $g^{(2)}_{\mathrm{auto}}$ and $g^{(2)}_{\mathrm{cross}}$ of coherent laser light. The dip in the auto-correlation function for values close to $\tau =0$ is caused by the pulse width of 1.5~ns of the detectors and the dead time of 2~ns of the time-to-digital converter used to record the signals. Aside from the dip and the auto-correlation peak, both functions show a constant value of one, as expected for a coherent source.}
	\label{fig:coherent_source_correlation}
\end{figure}

Our second experiment employs coherent states as the light source. The experimental setup and data analysis are, in principle, identical to those used for the single-photon source. In this case, the light source is a frequency- and power-stabilized 369.5~nm laser. It is referenced to an ultra-stable cavity, yielding a laser linewidth $1/T_\text{coh}$ of approximately 10~Hz. Light from the laser is scattered off the ion-trap assembly into the detection path, where the corresponding correlation functions are recorded. For this experiment, the power of the laser was increased suitably to overcome the ``losses'' introduced by the filter telescope in the detection path. The resulting measurements are shown in Fig.~\ref{fig:coherent_source_correlation}.
 
Both the PMTs and the time-to-digital converter have finite dead times of 1.5~ns and 2~ns, respectively. This instrumental limitation produces the dip observed in $g^{(2)}_{\mathrm{auto}}(\tau)$ in the time bins neighboring zero delay. In contrast, the cross-correlation is unaffected, as it involves two independent detectors. Aside from this effect, $g^{(2)}_{\mathrm{auto}}(\tau)$ and $g^{(2)}_{\mathrm{cross}}(\tau)$ are flat and equal to unity, as expected for a coherent source, except for the auto-correlation peak at $\tau=0$.

The average number of detected photon counts per binwidth $T_\text{bw} = 1$~ns is measured to be $\langle N\rangle = (1.19 \pm 0.01) \times 10^{-3}$, as extracted from the mean count rate of the detectors. Losses in the detection setup are already accounted for when using the detected count rate to calculate the mean photon number. For the auto-correlation, we find $g_\text{auto}^{(2)}(0) = 841.68 \pm 0.24$, which is in good agreement with the expectation of $g_\text{auto}^{(2)}(0) = 1 + 1/ \langle N\rangle$ for a coherent state. For the cross-correlation, we find a value of $g_\text{cross}^{(2)}(0) = 0.97 \pm 0.02$ close to one, as expected for a coherent source. Calculating the commutator using these values then yields $\langle\phi\vert [\hat{a},\hat{a}^{\dagger}]\vert\phi\rangle = 1.00 \pm 0.02$, which is again consistent with the theoretically expected value of one.

For $\tau \neq 0$, a straightforward extension of the arguments leading to \eqref{eq:ccr}  yields
\begin{equation}
	g^{(2)}_{\mathrm{auto}}(\tau) - g^{(2)}_{\mathrm{cross}}(\tau) = \frac{\langle  
		\{\hat{a}^{\dagger} (t) \hat{a} (t)\hat{a}^{\dagger} (t^{\prime})-\hat{a}^{\dagger} (t^{\prime})\hat{a}^{\dagger} (t)\hat{a}(t)\}\hat{a} (t^{\prime})
		\rangle}{ \langle \hat{a}^{\dagger} (t) \hat{a} (t) \rangle \langle \hat{a}^{\dagger} (t^{\prime}) \hat{a} (t^{\prime}) \rangle  \, } \, .
\end{equation}
From the  similarity of the two different experimental curves  for $\tau \neq 0$ we therefore conclude  that $[\hat{a}^{\dagger}(t^{\prime}), \hat{a}( t)] = 0$ and $[\hat{a}^{\dagger}(t^{\prime}), \hat{a}^{\dagger}( t)] = 0$. In combination with Eq.~\eqref{eq:ccr} this yields $[\hat{a}(t), \hat{a}^{\dagger}( t^{\prime})] = \delta (t- t^{\prime})$, implying that the field operators commute at different times~\cite{Blow:1990aa}. 

\textit{Concluding remarks.---}
We emphasize that the difference of  the auto- and cross-correlation functions of the light intensity contain useful information that is often overlooked. For instance, this difference can be used to improve the characterization of detector parameters , as discussed elsewhere. 
Here, we focus on the remarkable fact that the difference between the correlation functions at zero delay is directly proportional to the commutator of the quantum creation and annihilation operators. And conversely, the same experimental results show that for $t\neq t'$ the field operators commute.

Experimentally, we confirmed the commutator of these operators by measuring both the normalized auto- and cross-correlation functions for two fundamentally different light sources: a single-photon source and a coherent source. In both cases, the light was split by a beam splitter onto two detectors. The cross-correlation was obtained by correlating the independent detector time series, while for the auto-correlation, the two time series were first merged before correlating the resulting stream with itself. Using the same underlying data set for both correlations, we calculated the difference between the normally ordered cross-correlation and the not normally-ordered auto-correlation, and multiplied by the mean photon number. This procedure yielded a commutator value close to unity for both sources, in excellent agreement with theoretical expectations.

\emph{Acknowledgments.---} 
We acknowledge helpful discussions with  Gunnar Björk and Marco Bellini .  L. L. S.-S. was supported by the Spanish Agencia Estatal de Investigaci\'on (Grant PID2021-127781NB-I00). 

%

\end{document}